\documentstyle[11pt]{article}

\begin{document}

\setlength{\baselineskip}{0.8 cm}

\begin{center}
{\Large \bf On the Necessity of Rational Velocities: a Gedankenexperiment
\footnote{Work partially
supported by CNPq and FAPESP}}
\end{center}

\begin{center}
{\large \bf M.C. Nemes$^{(1)}$ and
Saulo C.S. Silva$^{(2)}$}
\end{center}

\begin{center}
$^{(1)}$Instituto de Ci\^encias Exatas, Universidade Federal de Minas
Gerais\\
CP702, 30000,
Belo Horizonte, MG, Brasil.
\end{center}

\begin{center}
$^{(2)}$Instituto de F\'{\i}sica, Universidade de S\~ao Paulo\\
CP20516, 01498, S\~ao Paulo, SP, Brasil.
\end{center}

\begin{abstract}
In the present contribution we propose a gedankenexperiment in which the
restriction of rational values on the velocities emerges as a necessary
condition from Classical Electromagnetism and Quantum Mechanics. This
restriction is shown to be intimately connected to Dirac's electric charge
quantization condition.
\end{abstract}

Important experimental results or currently accepted theoretical principles
remain unaffected if velocities are required to be rational numbers. This
interesting result has been recently given by Horzela et al$^{[1]}$. They
show, moreover, that such restriction can be easily incorporated into special
relativity, since rational velocities are Lorentz invariant quantities.

In the present contribution we propose a gedankenexperiment in which such
restriction on the velocities emerge as a necessary condition from Classical
Electromagnetism and Quantum Mechanics, intimately connected with Dirac's
electric charge quantization condition.

Let us consider the scattering of a charge $e$ under the influence of the
field generated by the charge distribution depicted in fig 1: two concurring
planes
with opposite charge densities, the concurring line being the $z$-axis. The
surface charge densities on the planes, $\sigma$, is given by

\begin{equation}
\sigma = \frac{g\rho}{4(\rho^{2}+z^{2})^{\frac{3}{2}}\theta}
\end{equation}

\noindent
where $g$ is a constant, $\theta$ is the angle between the concurring planes
and $\rho$ is the polar distance to the $z$-axis. The electric field generated
by such distribution (in the outside region, i.e., with polar angles $\phi$
outside the interval $[0,\theta]$) is given by

\begin{equation}
\vec{E} = \frac{g}{4\pi} \frac{\rho}{(\rho^{2}+z^{2})^{\frac{3}{2}}}
\hat{e}_{\phi}
\end{equation}

Now, as to the scattering situation, we assume that the charge $e$ is moving
with constant velocity $v$ in $z$-direction at a large enough impact parameter
$\rho=b$ such that the initial particle's trajectory remains unaffected.

The variation of the charge's momentum during the scattering can be
immediately calculated to be

\begin{equation}
\Delta\vec{p}=\int_{-\infty}^{+\infty}\vec{F}dt=\int_{-\infty}^{+\infty}
e\vec{E}dt=\int_{-\infty}^{+\infty}
\frac{eg}{4\pi}\frac{\rho}{(\rho^{2}+z^{2})^{\frac{3}{2}}}\hat{e}_{\phi}dt
\end{equation}

Inserting $\rho=b$ and $z=vt$ in $(3)$, we get

\begin{equation}
\Delta\vec{p}=\frac{egb}{4\pi}\hat{e}_{\phi}\int_{-\infty}^{+\infty}
\frac{dt}{(b^{2}+v^{2}t^{2})^{\frac{3}{2}}}=
\frac{egb}{2\pi v}\hat{e}_{\phi}
\end{equation}

To this momentum variation there will be a corresponding angular momentum's
change given by

\begin{equation}
\Delta\vec{L}=\frac{eg}{2\pi v}\hat{e}_{z}
\end{equation}

\noindent
which is independent of the impact parameter $b$. Using now Bohr's quantization
rule, we arrive at

\begin{equation}
\Delta L = \frac{eg}{2\pi v} = n
\end{equation}

The above relation can only be satisfied if we simultaneously fulfill

\begin{equation}
\frac{eg}{2\pi}=m
\end{equation}

\noindent
and

\begin{equation}
v=\frac{m}{n}
\end{equation}

The first one of these conditions is the celebrated Dirac's condition for
charge quantization$^{[2]}$, which restricts the values of $e$ and $g$ to
integer multiples of $e_{0}$ and $g_{0}=2\pi/e_{0}$ respectively. The
second condition represents the restriction of velocity values to rational
numbers. Since the charge's velocity is defined asymptotically, the condition
obtained cannot be regarded as an artifact of the particular scattering
situation. So we conclude that it arises as a natural consequence of
fundamental laws of Electromagnetism (Lorentz force equation) and Quantum
Mechanics (Bohr's quantization condition).

It is interesting to note at this point the connection between conditions
$(7)$ and $(8)$ and magnetic monopoles. The very same  conditions arise if
we were to consider the scattering of an electric charge $e$ by an axial
monopole$^{[3]}$ $g$ at rest since the field generated by such monopole is
precisely the same as the one given by $(2)$. This shows that the axial
monopoles proposed in $[3]$ can be interpreted as the particular charge
distribution considered here. If we take the limit $\theta\rightarrow 0$
and $\sigma\rightarrow\infty$ with

\begin{equation}
\sigma\theta=\frac{g\rho}{4(\rho^{2}+z^{2})^{\frac{3}{2}}}
\end{equation}

\noindent
remaining finite for $\rho$ and $z$ fixed, we would obtain a two-dimensional
topological defect, whose corresponding field is the axial monopole's field,
also given by $(2)$. Such physical picture of axial monopole as a domain-wall
is the subject of a forthcoming publication.

In spite of this interesting connection with axial magnetic monopoles, we
should like to stress that the results presented here, namely equations
$(7)$ and $(8)$, are independent of this picture.

\end{document}